\documentstyle[twocolumn,epsf]{jpsj}
\def\runtitle{ Field-Induced Gaps in the Frustrated Spin Ladder }

\title{Field-Induced Gaps in the Frustrated Spin Ladder
}
\author{ Nobuhisa {\sc Okazaki}, Kiyomi {\sc Okamoto}$^1$ and
T\^oru {\sc Sakai}}
\inst{Faculty of Science, Himeji Institute of Technology,\\Kamigouri-cho,
Akou-gun, Hyogo 678-1297, Japan \\ 
$^1$Department of Physics, Tokyo Institute of Technology,\\
Oh-okayama, Meguro-ku, Tokyo 152-0033, Japan}
\recdate{\today}
\abst{
We study the magnetization process of the $S=1/2$ antiferromagnetic spin ladder
in the presence of the second and the third-neighbor couplings
which lead to frustration with the typical nearest-neighbor coupling. 
We use degenerate perturbation theory and
level spectroscopy analysis of the numerical diagonalization
data of the Hamiltonian for finite systems.
We find two kinds of plateaux at half the saturation moment in the magnetization curve.
One is mainly due to the second-neighbor couplings
and the other to the third-neighbor couplings.
The mechanisms of these two plateaux are quite different with each other.
}

\kword{spin ladder, magnetization plateau, frustration}
\begin{document}
\sloppy
\maketitle

The spin gap is a current topic of interest in strongly correlated electron systems
because it is related to various interesting quantum phenomena
such as high-$T_c$ superconductivity.
Since the Lieb-Schultz-Mattis (LSM) theorem\cite{lsm} was recently generalized to
the magnetization process\cite{oshikawa},
a field-induced spin gap has attracted a great deal of interest
in the field of low-dimensional magnets.
The extended LSM theorem predicts that a 1D quantum spin system possibly
has gaps which are observed as plateaux in the magnetization curve
under the condition of quantization of the magnetization, described as
\begin{eqnarray}
Q(S-m) = {\rm integer},
\label{quantization}
\end{eqnarray}
where $Q$ is the spatial period of the ground state measured by the unit cell. 
$S$ and $m$ are the total spin and the magnetization per unit cell, respectively. 
Applying this theorem to the spin ladder, only the well-known spin gap
\cite{hida,dagotto,troyer} is expected to appear at $m=0$, as far as $Q=1$. 
Several theoretical analyses, however, predicted that field-induced gaps
would also appear at a finite magnetization,
with some modifications in the structure of the unit cell
such as three-leg\cite{cabra1} and bond-alternating ladders.\cite{cabra2} 

On the other hand, spontaneous breaking of the translational symmetry ($Q \ge 2$)
can also yield magnetization plateaux. 
The previous size scaling study\cite{okazaki} based on conformal field theory
indicated the possibility of the plateau at $m=1/2$ due to the two-fold degeneracy
of the ground state (i.e., $Q=2$) in the standard spin ladder
with the second-neighbor (2-N) interaction $J_2$. 
In the plateau phase, the singlet pair state and the $|\uparrow \uparrow \rangle $
state of the rung are expected to locate alternately along the leg. 
A similar mechanism of the field-induced gap was predicted for the zigzag
ladder equivalent to the bond-alternating chain with the 2-N interaction.
\cite{tonegawa,totsuka,tonegawa-okamoto}

In this paper, we investigate another mechanism of the plateau formation due
to the introduction of the third-neighbor (3-N) coupling $J_3$. 
We also present a typical phase diagram of the $J_3$-$J_2$ plane,
obtained by the level spectroscopy method
analyzing the finite cluster diagonalization data.

We consider the $S$=1/2 antiferromagnetic spin ladder with
2-N and 3-N exchange interactions in a magnetic field
described by the Heisenberg Hamiltonian
\begin{eqnarray}
{\hat H}&=&{\hat H}_0+{\hat H}_Z \\
{\hat H}_0&=&J_1\sum_i^L({\mib S}_{1,i} \cdot {\mib S}_{1,i+1}
+{\mib S}_{2,i} \cdot {\mib S}_{2,i+1}) \nonumber \\
& &+J_{\perp}\sum_i^L {\mib S}_{1,i} \cdot {\mib S}_{2,i} \nonumber\\
& &+J_2\sum_i^L({\mib S}_{1,i}
\cdot {\mib S}_{2,i+1}+{\mib S}_{2,i} \cdot {\mib
S}_{1,i+1})\nonumber\\
& &+J_3\sum_i^L({\mib S}_{1,i}
\cdot {\mib S}_{1,i+2}+{\mib S}_{2,i} \cdot {\mib
S}_{2,i+2})\\
{\hat H}_Z&=&-H\sum_i^L({S_{1,i}^z}+{S_{2,i}^z}),
\label{ham}
\end{eqnarray}
under the periodic boundary condition,
where $J_{1}$, $J_{\perp}$, $J_2$ and $J_3$ denote the coupling 
constants of the leg, rung and 2-N (diagonal) 
and 3-N exchange interactions, respectively (Fig. 1). 
\begin{figure}
\begin{center}
\leavevmode
\epsfxsize=7.0cm
\epsfysize=3.0cm
\epsfbox{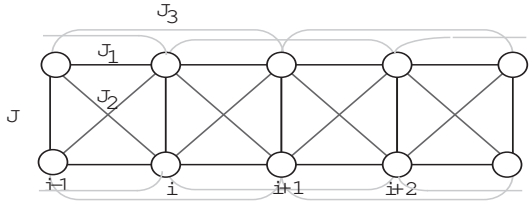}
\caption{Spin ladder with 2-N and 3-N exchange interactions
along the diagonals.
}
\label{fig.1}
\end{center}
\end{figure}
Hereafter we put $J_{\perp}$=1.
${\cal H}_Z$ is the Zeeman term where $H$ denotes the magnetic field
along the $z$-axis and the eigenvalue $M$ of the conserved quantity $\sum_{i}{({S_{1,i}^z}+
{S_{2,i}^z)}}$ is a good quantum number.
The macroscopic magnetization is represented by $m=M/L$.

In order to consider the possibility and the mechanism of the magnetization plateau at $m=1/2$
we use the degenerate perturbation theory around the strong rung coupling limit
$J_1,J_2,J_3 \ll 1$.\cite{mila} 
We introduce a pseudo spin $\mib T$ for each rung pair and map the two original states
$(|\uparrow \downarrow \rangle -|\downarrow \uparrow \rangle )/\sqrt{2}$
and $|\uparrow \uparrow \rangle $ of the $\mib S$ picture to the
$|\Downarrow \rangle $ and $|\Uparrow \rangle $ states of $\mib T$, 
neglecting the other two states
$(|\uparrow \downarrow \rangle + |\downarrow \uparrow \rangle )/\sqrt{2}$
and $|\downarrow \downarrow \rangle $. 
After the mapping, 
we obtain the effective Hamiltonian
\begin{eqnarray}
{\hat H}_{\rm eff}=
(J_1-J_2)\sum_i^L({T}^x_{i} \cdot {T}^x_{i+1}
+{T}^y_{i} \cdot {T}^y_{i+1})\nonumber\\
+\frac{J_1+J_2}{2} \sum_i^L({T}^z_{i} \cdot {T}^z_{i+1}) \nonumber\\
+J_3\sum_i^L({T}^x_{i} \cdot {T}^x_{i+2}
+{T}^y_{i} \cdot {T}^y_{i+2})\nonumber\\
+\frac{J_3}{2}\sum_i^L({T}^z_{i} \cdot {T}^z_{i+2}).
\label{eham}
\end{eqnarray}
This effective Hamiltonian describes the $T=1/2$ $XXZ$ chain with 2-N interactions, 
where $J_2$ and $J_3$ control the $XXZ$ anisotropies and 2-N couplings, respectively, 
with fixed $J_1$. 
We note that the $XXZ$ anisotropy parameters of the NN and 2-N interactions
are different from each other.
Well-established works on this model have revealed the following properties: \cite{no}
the system has three phases; the spin fluid (gapless), N\'eel (gapful) and dimer (gapful) phases. 
The $J_2=J_3=0$ case is clearly in the gapless phase and sufficiently large $J_2$ ($J_3$)
yields the N\'eel (dimer) phase via the Berezinski-Kosterlitz-Thouless 
(BKT) transition.\cite{berezinski,kt}
The boundary between the N\'eel and dimer phases is the Gaussian line.
The above properties lead us to the conclusion that $J_2$ and $J_3$ give rise to the plateau
at $m=1/2$ in the original system,
based on different mechanisms; the N\'eel state and the dimer state, in the language of pseudo spins. 
The two plateaux are hereafter called the plateau A (N\'eel) and plateau B (dimer). 
Clearly, both plateau phases should be accompanied by the two-fold degeneracy
due to the spontaneous breaking of the translational symmetry.

Next, we perform a numerical analysis for the original Hamiltonian (\ref {ham})
to more quantitatively confirm the realization of  two plateaux at $m=1/2$,
predicted by the degenerate perturbation theory. 
The {\it level spectroscopy}\cite{no} is
a powerful method to determine the BKT boundary,
as well as most second-order transitions in 1D quantum systems.
In this method, the phase boundaries can be determined from the level crosses
between low-lying excitations.
This method is free from the most dominant logarithmic size corrections,
which make it difficult to determine the BKT boundaries when conventional methods are applied.
Hereafter, $E(L,M,k)$ indicates the lowest eigenvalue 
of the Hamiltonian ${\hat H}_0$ in the subspace
where the eigenvalue of $\sum_{i}{({S_{1,i}^z}+{S_{2,i}^z)}}$ is $M$
and the momentum is $k$ for the system size $L$.
Using the Lanczos algorithm, $E(L,M,k)$ is calculated for $L=4n$ $(\leq 16)$,
to avoid frustration among the 3-N exchange interactions under the periodic boundary condition. 
Before investigating the full Hamiltonian,
let us briefly demonstrate the powerfulness of level spectroscopy
by drawing a phase diagram for the $J_3=0$ case,
which has already been obtained by conventional methods.\cite{okazaki}
In level spectroscopy,
we use the following two excitations given by
\begin{eqnarray}
\Delta _1
&=&\frac{1}{2}\left\{E\left(L,{\frac{L}{2}}+1,\pi\right)+E\left(L,{\frac{L}{2}}-1,\pi\right) \right\}
 \nonumber \\
& &-E\left(L,{\frac{L}{2}},0\right), 
\label{gap1}
\end{eqnarray}
and
\begin{eqnarray}
\Delta _0
=E\left(L,{\frac{L}{2}},\pi\right)-E\left(L,{\frac{L}{2}},0\right).
\label{gap0}
\end{eqnarray}
When the parameters are swept, the state is gapless or gapful (i.e., plateau)
according to whether $\Delta_1 < \Delta_0$ or $\Delta_1 > \Delta_0$,
as explained by Okamoto et al.\cite{oka-tone}
The phase diagram of the $J_2$-$J_1$ plane, obtained by this procedure is shown in Fig. 2. 
\begin{figure}
\begin{center}
\leavevmode
\epsfxsize=7.0cm
\epsfysize=6.0cm
\epsfbox{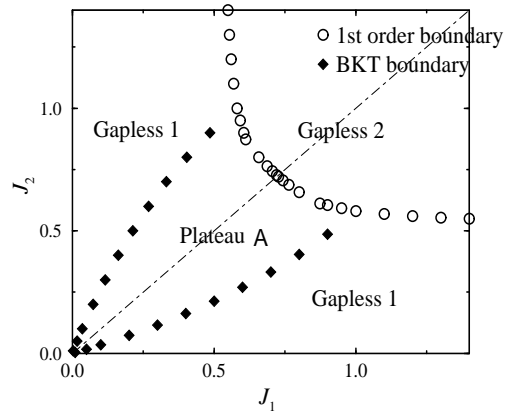}
\caption{
Phase diagram on the $J_2$-$J_1$ plane at $m=1/2$.
}
\label{fig.2}
\end{center}
\end{figure}
The estimated error bars of the boundary points are much smaller than the size of the marks.
In this phase diagram, in addition to the original spin fluid (gapless 1) phase,  
there appears another gapless phase (gapless 2). 
Here we do not touch the gapless 2 phase (equivalent to the $S=1$ chain at $m=1/2$;
see refs. 8 and 17 for details).
Using the effective Hamiltonian eq. (5),
we see that the slope of the boundary line between the plateau A and gapless 1 phases
is $1/3$ in the limit of $J_1\to 0$
(note that $H_{\rm eff}$ of eq. (5) is exact in this limit),
as noted by Mila.\cite{mila}.
In our numerical calculation, we obtain $J_2/J_1 = 0.3350$ when $J_1=0.01$,
which agrees very well with the exact value $1/3$.
This shows the high reliability of our level spectroscopy method.
When conventional methods are applied to this problem,\cite{so,okazaki}
the slope value in the $J_1 \to 0$ limit is estimated as $J_2/J_1 \simeq 0.45$,
which is much larger than the exact value $1/3$.
The reason for this difference is discussed by Okamoto\cite{okamoto} in detail
and will be published elsewhere.

\begin{figure}
\begin{center}
\leavevmode
\epsfxsize=7.0cm
\epsfysize=6.0cm
\epsfbox{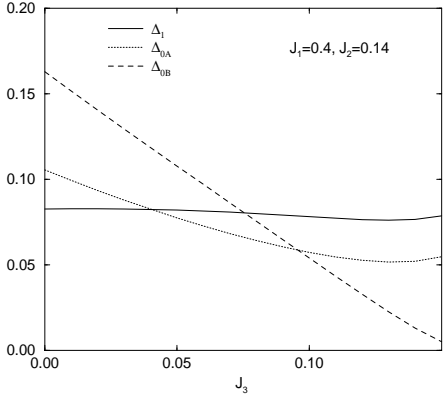}
\caption{
Excitations $\Delta_{0\rm A}$, $\Delta_{0\rm B}$ and $\Delta_1$.
}
\label{fig.3}
\end{center}
\end{figure}
Here we consider the full Hamiltonian problem $J_3 \ne 0$. 
For simplicity, we fix the 2-N interaction as $J_1=0.4$. 
The gapless 2 phase does not appear in this case. 
Here, we consider two excitations having $M=L/2$ and $k=\pi$,
because the change in the lowest energy level in this sector occurs when $J_3$ is increased.
Then, we define $E_A(L,{L\over 2},\pi)$ ($E_B(L,{L\over 2},\pi)$) as the lower
energy
in the region of large $J_2$ ($J_3$) and small $J_3$ ($J_2$),
and also define $\Delta _{0A}$ ($\Delta _{0B}$) in the same way as the eq. (\ref {gap0}).
These two excitations $E_A(L,{L\over 2},\pi)$ and $E_B(L,{L\over 2},\pi)$ correspond to
the N\'eel and dimer excitations (see ref. 13) in the picture of the pseudo spin
$\mib T$,
and can be distinguished by the eigenvalues $P$ of the space inversion operation
of the rung number $i \to L-i+1$.
Namely, when $L=4n$, the state for $E_A(L,{L\over 2},\pi)$ has $P=-1$ and that for
$E_B(L,{L\over 2},\pi)$ has $P=+1$.
In the spin fluid (gapless 1), plateau A and plateau B phases, 
the lowest excitations should be $\Delta _1$, $\Delta _{0A}$ and $\Delta _{0B}$, respectively. 
Therefore, the phase boundaries can be determined from the level crossing points
among these three excitations.\cite{no}
Figure 3 shows these three excitations as functions of $J_3$ when $J_1=0.4, J_2=0.14, L=12$.
Thus, the state is gapless for $J_3 <0.040$, plateau A for $0.040 < J_3 < 0.095$,
and plateau B for $J_3 > 0.095$.
Repeating such procedures with sweeping the parameters,
we obtain the phase diagram of the $J_2-J_3$ plane as shown in Fig. 4. 
\begin{figure}
\begin{center}
\leavevmode
\epsfxsize=7.0cm
\epsfysize=6.0cm
\epsfbox{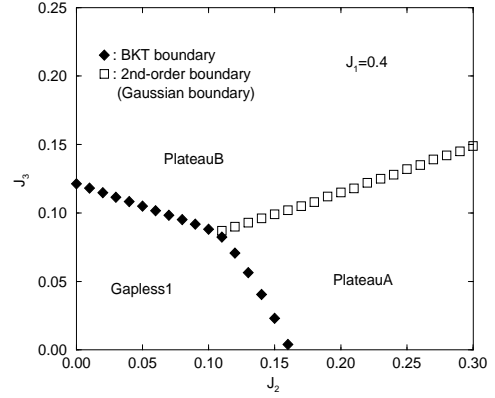}
\caption{
Phase diagram of the $J_2$-$J_3$ plane with fixed $J_1$ (=0.4) at
$m=1/2$.
}
\label{fig.4}
\end{center}
\end{figure}
\begin{figure}
\begin{center}
\leavevmode
\epsfxsize=7.0cm
\epsfysize=6.0cm
\epsfbox{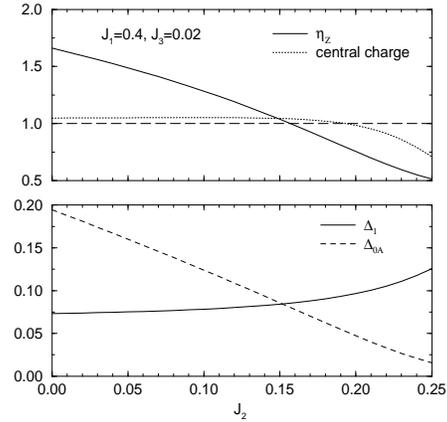}
\caption{
Central charge $c$ and critical exponent $\eta_z$ near the BKT boundary
between the gapless and plateau A phases,
as well as excitations $\Delta_1$ and $\Delta_{0\rm A}$.
}
\label{fig.5}
\end{center}
\end{figure}
In order to investigate the universality of the boundary between the gapless and plateau A phases, 
we estimate the central charge $c$ by\cite{cft}
\begin{eqnarray}
\frac{1}{L}E\left(L,{L\over 2},0\right)
={\epsilon_{\infty}}- \frac{\pi}{6}cv_{\rm s}
\frac{1}{L^2} \hspace{1cm} (L\to\infty).
\label{central}
\end{eqnarray}
where $v_{\rm s}$ is the sound velocity, which can be calculated by 
\begin{eqnarray}
v_{\rm s}=
\frac{L}{2\pi}\left[E\left(L,{L\over 2},{{2\pi}\over L}\right)
-E\left(L,{L\over 2},0\right)\right],  (L\to\infty).
\end{eqnarray}
We also estimate the critical exponent $\eta_z$ defined by 
$\langle S_0^z S_r^z \rangle - \langle S^z \rangle^2 \sim (-1)^r r^{-\eta_z}$ by use of\cite{no}
\begin{equation}
    \eta_z
    = {L \over 2\pi v}(3\Delta_{0\rm A} + \Delta_{0\rm B})
    \label{etaz}
\end{equation}
near the gapless 1-plateau A boundary.
We note that the roles of $\Delta_{0\rm A}$ and $\Delta_{0\rm B}$ are interchanged 
near the  gapless 1-plateau B boundary.
Since the most dominant logarithmic size corrections are canceled out in eq. (\ref {etaz}),\cite{no}
we can obtain an accurate value of $\eta_z$ from eq. (\ref {etaz}).
At the BKT transition point of the present type,
the exponent $\eta_z$ should be unity.
Figure 5 shows the behaviors of $c$ and $\eta_z$ near the gapless 1-plateau A boundary,
as well as the level cross between $\Delta_1$ and $\Delta_{0\rm A}$,
resulting in  $J_2^{\rm (cr)} \simeq 0.15$.
We can clearly see that $\eta_z \simeq 1$ at $J_2^{\rm (cr)}$,
which strongly confirms the universality class of the BKT transition.
The behavior of $c$ also suggests the BKT transition.
A similar conclusion is also obtained for the gapless 1-plateau B transition.

To clarify the properties of the boundary between plateaux A and B, 
we show the $J_3$ dependence of the scaled gap $2L\Delta _1$
(2$\Delta _1$ is the length of the plateau of system size $L$)
along the line $J_2=0.3$ in Fig. 4. 
\begin{figure}
\begin{center}
\leavevmode
\epsfxsize=8.0cm
\epsfysize=6.0cm
\epsfbox{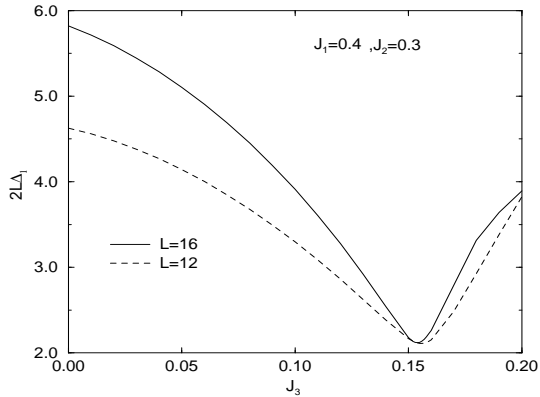}
\caption{Behavior of the scaled gap $2L\Delta _1$ in the $J_1$=0.4 and $J_2=0.3$ case.
}
\label{fig.6}
\end{center}
\end{figure}
The size dependence of the scaled gap suggests that the plateau is opening
in both phases and that the system is gapless only at the boundary $J_3 \simeq 0.15$. 
We also found $c=1$ on the line labeled by open squares in Fig. 4. 
These results are consistent with the Gaussian fixed line
predicted by degenerate perturbation theory. 

Several magnetization curves are also presented on line $J_2=0.3$;
$J_3=$0, 0.10, 0.15 and 0.20. 
They are obtained by the size scaling analysis\cite{st} based on conformal field theory
and Shanks transformation\cite{shanks}, using the numerical result of $E(L,M,k)$ up to $L=16$. 
Only the lines of fitting polynomials are shown in Fig. 7. 
\begin{figure}
\begin{center}
\leavevmode
\epsfxsize=8.0cm
\epsfysize=6.0cm
\epsfbox{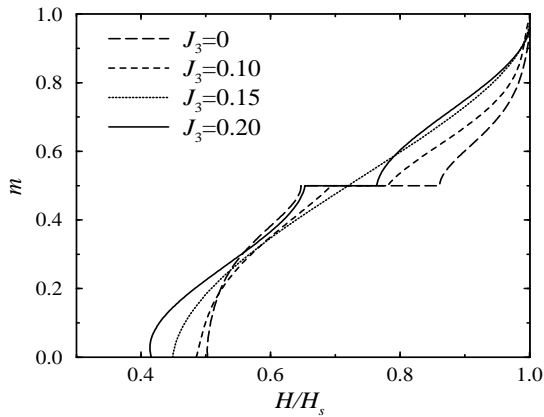}
\caption{Magnetization curves for $J_1=0.4$ and $J_2$=0.3, with various
$J_3$
(=0, 0.05, 0.1, 0.15 and 0.2).
}
\label{fig.7}
\end{center}
\end{figure}
This suggests that, 
with increasing $J_3$, the plateau decreases until vanishing at the critical point
and then increases again. 
This behavior also explains why the mechanism of the gap formation due to $J_3$
is different from the one due to $J_2$. 

According to the present analysis, the gapless-plateau critical
value of $J_3$ for $J_2=0$ is smaller than that of $J_2$ for $J_3=0$, irrespective of $J_1$. 
In addition, the plateau B phase can appear for any ratio of $J_1/J_{\perp}$,
although the plateau A phase cannot appear for $J_1/J_{\perp} > 1$. 
Thus, the $J_3$-induced plateau is more realistic than the one due to $J_2$. 
In fact, a typical spin ladder, $\rm{SrCu_2O_3}$\cite{ladder}, 
was reported to hold $J_1/J_{\perp} \sim 2$. 
Thus, a  plateau might be caused by the 3-N coupling of $J_3$ in this or related materials.

In summary, the magnetization process of the frustrated spin ladder
was investigated with degenerate perturbation theory and the level spectroscopy. 
The present analysis revealed the appearance of a novel magnetization plateau at $m=1/2$
due to the 3-N interaction. 
The mechanism of this plateau is explained by the spontaneous dimerization
of the pseudo spin system. 
As far as we know, this is the first theoretical finding of the change in the
plateau mechanisms, both of which have the spontaneous symmetry breaking, 
by sweeping physically natural parameters.


\section*{References}
\begin{thebibliography}{99}
\bibitem{lsm}
   E. Lieb, T. D. Schultz and D. C. Mattis:
   Ann. Phys. (N. Y.) {\bf 16} (1961) 407.
\bibitem{oshikawa}
   M. Oshikawa, M. Yamanaka, and I. Affleck:
   Phys.Rev.Lett. {\bf 78} (1997) 1984.
\bibitem{hida}
   K. Hida:
   J. Phys. Soc. Jpn. {\bf 60} (1991) 1347 .
\bibitem{dagotto}
   E. Dagotto, J. Riera and D. Scalapino:
   Phys. Rev. B {\bf 45} (1992) 5744.
\bibitem{troyer}
   M. Troyer, H. Tsunetsugu and T. M. Rice:
   Phys. Rev. B {\bf 53} (1996) 251.
\bibitem{cabra1}
   D. C. Cabra, A. Honecker and P. Pujol: 
   Phys. Rev. Lett. {\bf 79} (1997) 5126. 
\bibitem{cabra2}
   D. C. Cabra and M. D. Grynberg:
   Phys.Rev.Lett.{\bf 82} (1999) 1768.
\bibitem{okazaki}
   N. Okazaki, J. Miyoshi and T. Sakai:
   J. Phys. Soc. Jpn. {\bf 69} (2000) 37.
\bibitem{totsuka}
   K. Totsuka:
   Phys.Rev.{\bf B57} (1998) 3435.
\bibitem{tonegawa} T. Tonegawa, {\it et al.}:
   Physica. {\bf B246-247} (1998) 509.
\bibitem{tonegawa-okamoto}
  T. Tonegawa, K. Okamoto and M. Kaburagi:
  in preparation
\bibitem{mila}
  F. Mila:
  Eur. Phys. J. {\bf B6} (1998) 201.
\bibitem{no} 
  K. Nomura and K. Okamoto:
  J. Phys. A: Math. Gen. {\bf 27} (1994) 5773.
\bibitem{berezinski}
  V. L. Berezinski:
  Zh. Eksp. Teor. Fiz. {\bf 59} (1970) 907;
  Sov. Phys. JETP {\bf 32} (1971) 493.
\bibitem{kt}
  J. M. Kosterlitz and D. J. Thouless:
  J.Phys.C {\bf 6} (1973) 1181
\bibitem{oka-tone}
  K. Okamoto, T. Tonegawa, Y. Takahashi and M. Kaburagi:
  J. Phys. A: Math. Gen {\bf 11} (1999) 10485.
\bibitem{so} 
  T. Sakai and N. Okazaki:
  J. Appl. Phys. to appear.
\bibitem{okamoto}
  K. Okamoto: in preparation.
\bibitem{cft}
  J. L. Cardy:
  J .Phys. A {\bf 17} (1984) L385(1984);
  H.W.Bl\"{o}te, J. L. Cardy and M. P. Nightingale:
  Phys. Rev. Lett. {\bf 56} (1986) 742;
  I.Affleck, {\it ibid}.:{\bf 746} (1986).
\bibitem{st} 
  T. Sakai and M. Takahashi:
  Phys.Rev.B {\bf 57} (1998) R3201.
\bibitem{shanks}
  D. Shanks: 
  J. Math. Phys. {\bf 34} (1955) 1.
\bibitem{ladder}
  M. Azuma, {\it et al.}:
  Phys. Rev. Lett {\bf 73} (1994) 3463.
\end{thebibliography}
\end{document}